\documentclass[submission,copyright,creativecommons]{eptcs}
\providecommand{\event}{TAV-WEB 2010} % Name of the event you are submitting to
\usepackage{breakurl}             % Not needed if you use pdflatex only.

\usepackage{listings}
\usepackage{epsfig}
\title{Relational Constraint Driven Test Case Synthesis\\ for Web Applications}
\author{Xiang Fu
\institute{Hofstra University\\ Hempstead, NY 11549}
\email{Xiang.Fu@hofstra.edu}
}
\def\titlerunning{Relational Constraint Driven Test}
\def\authorrunning{X. Fu}
\begin{document}
\maketitle

% definitions

 \long\def \ignoreme#1{}
 \def\ttt#1{{\texttt{#1}}}
 \def\text#1{{\textrm{#1}}}
 \def\tb#1{{\textbf{#1}}}
 \def\caA{{\cal A}}
 \def\caP{{\cal P}}
 \def\caL{{\cal L}}
 \def\caB{{\cal B}}
 \def\caD{{\cal D}}
 \def\caC{{\cal C}}
 \def\caT{{\cal T}}
 \def\sender{\mbox{\sc sen}}
 \def\receiver{\mbox{\sc rec}}
 \def\Min{M^{in}}
 \def\Mout{M^{out}}
 \def\attr{\mbox{\sc attr}}
 \def\domain{\mbox{\sc dom}}
 \def\dom{\mbox{\sc dom}}
 \def\class{\mbox{\sc type}}
 \def\content{\mbox{\sc con}}
 \def\xpath{\mbox{\sc xpath}}
 \def\And{~\wedge~}
 \def\Or{~\vee~}
 \def\Imply{~\Rightarrow~}
 \def\wdom{\widehat{\domain}}
 \def\Sigin{\Sigma^{i}}
 \def\Sigout{\Sigma^{o}}
 \def\caP{{\cal P}}
 \def\caS{{\cal S}}
 \def\caR{{\cal R}}
 \def\joinc{\mbox{\sc joinc}}
 \def\join{\mbox{\sc join}}
 \def\intersect{~\cap ~}
 \def\union{~\cup ~}
 \def\reverse{\ttt{reverse}}
 \def\maxrho{\rho_{max}}
 \def\paragraph#1{\vspace*{0.1in} \noindent {\bf #1:} }
 \def\mit#1{{\it #1}}
 \def\mrm#1{{\rm #1}}
\def\skeleton{{\rm skeleton}}
\def \subs [#1][#2][#3]{{{#1}_{{#2} \rightarrow {#3}}}}
\def \reluctant [#1]{\ttt{reluc}(#1)}
\def \subsd [#1][#2][#3]{{{#1}_{{#2} \rightarrow {#3}}}}
\def \subsg [#1][#2][#3]{{{#1}^+_{{#2} \rightarrow {#3}}}}
\def \subsr [#1][#2][#3]{{{#1}^-_{{#2} \rightarrow {#3}}}}
\def \fst[#1][#2][#3]{{{\cal M}_{{#1} \rightarrow {#2} ~\Rightarrow {#3}}}}
\def \fstg[#1][#2][#3]{{{\cal M}^+_{{#1} \rightarrow {#2} ~\Rightarrow {#3}}}}
\def \fstr[#1][#2][#3]{{{\cal M}^-_{{#1} \rightarrow {#2} ~\Rightarrow {#3}}}}
\def \orgfst[#1][#2]{{{\cal M}_{{#1} \rightarrow {#2}}}}
\def \orgfstg[#1][#2]{{{\cal M}^+_{{#1} \rightarrow {#2}}}}
\def \orgfstr[#1][#2]{{{\cal M}^-_{{#1} \rightarrow {#2}}}}
\def\replace{\subs[E][r_1][w_2]}
\def\substring [#1][#2][#3]{{#1}[#2,#3]}
\def \varset [#1]{{\vec{#1}^v}}
\def \regset [#1]{{\vec{#1}^r}}
\def \reg [#1]{{\vec{#1}^r}}
\def \weakequiv{{\equiv_{\it w}}}
\def\qed{\hfill\rule{1mm}{1.5ex}}
\def \strongequiv{{\equiv_{\it s}}}
\def \sp{\ttt{sp}}
\def \null {\ttt{null} }
\def \SAT {\ttt{SAT} }
\def \Pre{\ttt{Pre} }
\def \Post{\ttt{Post} }
\def \PostTrans{\ttt{PostTrans} }

\def \dom {{\tt Dom} }
\def \D {{\cal D} }
\def \V {{\cal V} }
\def \A {{\cal A} }
\def \R {{\cal R} }

\newtheorem{itallemma}{Lemma}[section]
\newtheorem{italdeff}[itallemma]{Definition}
\newenvironment{definition}{\begin{italdeff}\em}{\qed\end{italdeff}}
\newenvironment{lemma}{\vspace*{2pt}\begin{itallemma}\rm}{%
   \end{itallemma}\vspace*{2pt}}
\newtheorem{equivalence}{Equivalence}
\newtheorem{italexamp}[itallemma]{Example}
\newenvironment{examp}{\begin{italexamp}\em}{\qed\end{italexamp}}
\newtheorem{italthm}[itallemma]{Theorem}
\newenvironment{thm}{\vspace*{2pt}\begin{italthm}\rm}{%
  \end{italthm}\vspace*{2pt}}

\begin{abstract}
%low coverage of test cases that are designed manually. 
This paper proposes a relational constraint driven technique
that synthesizes test cases automatically for web applications. 
Using a static analysis, servlets can be modeled as
relational transducers, which manipulate backend databases.
We present a synthesis algorithm that generates
a sequence of HTTP requests for simulating a user session.
The algorithm relies on backward symbolic image computation
for reaching a certain database state, given a code coverage objective.
With a slight adaptation, the technique can be used for discovering
workflow attacks on web applications.
\end{abstract}

\section{Introduction}
%1. problem and motivation
  Modern web applications usually rely on backend database systems
for storing important system information or supporting business decisions.
The complexity of database queries, however, often complicates the task of
thoroughly testing a web application. To manually design
test cases involves labor intensive initialization of database systems, 
even with the help of unit testing tools such as SQLUnit \cite{SQLUnit} and 
DBUnit \cite{DBUnit}. It is desirable to
automatically synthesize test cases for web applications.
 
%2. background introduction: web application testing
  There has been a strong interest recently in testing database driven 
applications and database management systems
(see e.g., \cite{Emmi07issta,Lo07icde,ADUSA}). 
Many of them are query aware, i.e.,
given a SQL query, an initial database (DB) instance is generated
to make that query satisfiable. The DB instance is fed
to the target web application as input, so that
a certain code coverage goal is achieved.
The problem we are trying to tackle is one step further -- it is
a {\em synthesis problem}: given a certain database state 
(or a relational constraint), a {\em call sequence} of web servlets is
synthesized to reach the given DB state.  This is
motivated by the special architecture of web applications.
Unlike typical desktop software systems, the atomic components
of a web application, i.e., {\em web servlets}, 
are accessible to end users.  In addition, cookies and session variables are 
frequently used for session maintenance and user tracing.
Thus web application testing is usually session oriented 
(see, e.g., \cite{Sampath03,Elbaum05tse}).  
Unlike unit testing general database driven applications,
an intermediate database state has to be resulted from a call sequence
and cannot be initialized at will. We are interested in synthesizing
such a call sequence. The technique could also be leveraged to
detect workflow attacks \cite{Balzarotti07CCS,Cova07RAID},
where an ``unexpected" call sequence can cause harm 
to the business logic of a web application (e.g., shipping a product
without charging credit card).

%3. our proposal
We propose a white-box analysis, which consists of the following steps:
(1)  {\em interface extraction}: each web servlet is modeled as a collection of
{\em path transducers}. A path transducer is essentially a relational
transducer \cite{AVFY00} that corresponds to a single
execution path of the web servlet. A transducer is represented
as a pair of pre/post conditions, built upon relational constraints.
Path transducers are extracted using light-weight symbolic execution
technique. (2) {\em coverage goal generation:} to achieve a certain
coverage goal, symbolic execution is used to generate relational
constraints (expressed using first order relational logic 
\cite{Jackson00FSE}).  (3) {\em call sequence synthesis:} 
a heuristic algorithm is used to determine
a call sequence that could lead to the required intermediate database state.
Symbolic constraint solving is currently used for performing
backward/forward symbolic image computation of transducers.
Best effort constraint solving (restricted to finite model and finite
scope), based on Alloy Analyzer \cite{Alloy}, is used to 
generate parameters of each HTTP request in the call sequence.  
% (4) {\em test case execution and evaluation:} 
% we then replay each test case and evaluate the effectiveness
% of the approach in practice. The implementation of the proposed
% framework is currently under development.

% %4. contributions
%   This paper has made the following contributions:
% \begin{enumerate}
%   \item A static analysis approach for interface extraction of web servlets
%   as relational transducers.
%   \item A high precision constraint solving technique and the 
%   symbolic backward/forward image computation technique for relational
%   constraints.
%   \item A query-aware call sequence synthesis algorithm that improves
%   code coverage of web application testing.
%   \item Application and evaluation of the techniques over a wide collection
%   of open-source web applications. 
% \end{enumerate}

%5. Paper organization
 Although there does not exist an implementation for the proposed
technique, we plan to illustrate its effectiveness and
feasibility using a case study.
\S{2} presents a motivating example.
\S{3} introduces the path transducer model. \S{4} discusses
symbolic image computation of relational algebra. The call sequence
algorithm is presented in \S{5}. \S{6} 
adapts the algorithm for discovering workflow attacks.
\S{7} discusses related work and concludes.

\section{Motivating Example}

This section introduces \ttt{SimpleScarf},
a case study example used throughout the paper.  
\ttt{SimpleScarf} is adapted from the 
Scarf conference management system \cite{scarf}.  It is comprised of 
five servlets for managing the membership of a conference.  
Each servlet is implemented as a PHP file that accepts HTML requests
and generates HTML responses.

These servlets are briefly described as follows. Later we will
present the formal relational transducer model for each of them. 
(1)
\ttt{Showsessions.php} displays the list of paper sessions that
are associated with the current user.
(2)
\ttt{Insertsession.php} adds a new paper session to the system. 
(3) \ttt{Addmember.php} inserts a new member into a paper session. 
(4) \ttt{Generaloptions.php} creates a new user of the system.
(5) \ttt{Login.php} authenticates the log-in process.
Once a user successfully logs in, the servlet sets 
a session variable called ``\ttt{user}" for tracking
the user session.

There is a backend database with three relations:
(1) $\ttt{Users}(\ttt{uname}, \ttt{enc\_wd})$ contains information
about a user: the user name and the encrypted password;  
(2) $\ttt{Sessions} (\ttt{sid}, \ttt{sname})$ has two attributes: the paper 
session id and session name;
and
(3) $\ttt{Member}(\ttt{sid}, \ttt{uname})$ describes the members
of a paper session.  To simplify the scenario, the data type of
each column is \ttt{varchar}.
For relation \ttt{Member},
there are two foreign key dependency constraints: (1) \ttt{sid} 
on \ttt{Sessions}, and (2) \ttt{uname} on \ttt{Users}.
Throughout the paper, each relation or servlet
is denoted using the first character
of its name, e.g., \ttt{Showsessions.php} denoted by \ttt{S}
and \ttt{Users} denoted by \ttt{U}. 

%Foreign key constraints:
%\begin{enumerate}
% \item \ttt{members-users} (C1): 
%  $
%	\pi_2(M) \subseteq \pi_1(U).
%  $
% \item \ttt{sessions-users} (C2): 
%  $
%	\pi_2(M) \subseteq \pi_1(U).
%  $
% \item \ttt{primary key of users} (C3): 
%  $
%	(x,y) \in U \And (x,y_2) \in U \Imply y = y_2
%  $
% \item \ttt{primary key of users} (C4): 
%  $
%	(x,y,z) \in S \And (x,y_2,z_2) \in S \Imply y = y_2 \And z=z_2
%  $
% \item \ttt{primary key of users} (C5): 
%  $
%	(x,y) \in M \And (x,y_2) \in M \Imply y = y_2 
%  $
%\end{enumerate}

As an example, Listing\ \ref{listing:showsession} presents a fragment of
\ttt{Showsessions.php}. The servlet consists of two parts:
(1) a branch statement which examines a user's login status,
 by checking the value of session variable ``\ttt{uname}"; 
and (2) a loop that reads the information retrieved
by a \ttt{SELECT} statement to generate 
a list of paper sessions related to the user.

\begin{figure}[h!]
\begin{center}
\lstset{language=PHP, frame=single,  caption = Fragment of Showsessions.php,
label=listing:showsession, basicstyle=\scriptsize, captionpos=b}
\begin{lstlisting}
1<?php
2 include_once("header.php");
3 print "<div style='float:right'>";
4 if (!isset($_SESSION['uname'])) 
5    print "<a ...>Not logged in</a><br> ";
6 else 
7    print "Logged in as " $_SESSION['uname'] ..
8 print "</div><br>";
12 $result = query("SELECT session_id ... ");
13 if (mysql_num_rows($result) == 0) {
14     print ("Sorry, ...") 
15 }
16 $curday = -1;
44 while ($row = mysql_fetch_array($result)) {
45    $day = date("F j", $row['starttime']);
   ...
90} 
\end{lstlisting}
\end{center}
\vspace*{-8mm}
\end{figure}

\paragraph{Coverage Goal}
 line \ttt{45} in Listing\ \ref{listing:showsession}. We are interested in
two questions: (1) what is the database state or relational constraint
that leads the execution of \ttt{Showsessions.php}
to line \ttt{45}? (2) what is the call sequence that generates
the desired database state?
Later, we will show that an algorithm can automatically synthesize 
a call sequence \ttt{IGALS} and the HTML parameters in each
HTTP request.

\section{Path Transducer Model}
 We intend to model each {\em execution path} of a servlet, as an
{\em atomic relational transducer}.  For example, 
one such execution path of 
\ttt{Showsessions.php} could be 
1-2-3-4-6-7-8-12-13-15-16-44-45-44-90 (line numbers).  
We trace such an execution using a ``path condition", i.e.,
the conjunction of branch conditions along the
execution path.  Note that each servlet may have
an infinite set of such transducers. During a static analysis,
we can bound loop iterations and recursive call depth
for achieving a finite set.

 Because each transducer only models one path, it is
essentially a transition rule that is expressed as a Boolean combination
of relational expressions. 
The first order relational logic \cite{AHV95} is
suitable for expressing uninterpreted functions and
next state variables and relations.
Here a primed variable (i.e., with a single 
quote) represents the value of
the variable in the next state. Input parameters are
preceded with a dollar sign, e.g., $\$u,\$p,\$s$ represent the
input parameters \ttt{uname}, \ttt{password}, \ttt{session name} in
a HTTP request.
Session variables are denoted
using $\#$, e.g., $\#u$ represents the session variable \ttt{uname}.
If in a transition rule,
a session variable or relation does not have the primed form, then its value
remains the same after the execution of the transition.  
Relations are generally represented as sets of tuples. 

\begin{definition}  \label{deff:transducer}
A path transducer is a 
tuple $(\caD, \dom, \vec{V}, \vec{A})$
where $\caD$ is the data schema (a finite set of relation schema),
\dom is an infinite but countable domain for $\caD$,
$\vec{V}$ is a finite set of session variables (letting
$\vec{V}'$ be the set of primed variables), and
$\vec{A}$ is a Boolean combination of terms. Each term has one
of the following forms:
\begin{enumerate}
  \item {\em (equality) $u = E(\vec{v},\dom)$}, 
where $u\in \vec{V} \cup \vec{V}'$
and $E$ is an expression on the current state of variables and
constants from $\dom$.
  \item (satisfiability check) $\SAT(\Psi)$ where $\Psi$ is 
a relational algebra formula on $\caD$.
\end{enumerate}
\end{definition}

 Notice that in the equality form, a primed variable (next state) is allowed
to appear on the LHS (left hand side) only. The syntax (expressiveness)
of $E$ is affected by the decision procedure used in analysis.
Currently, we allow Presburger arithmetic \cite{KMP95} and
Simple Linear String Equation \cite{nfm2010}. Uninterpreted functions
are allowed with first order relational logic on $\caD$, but not
with Presburger arithmetic.
A relational constraint is defined similarly as a transducer.
It is a Boolean combination of equality and satisfiability terms,
except that no primed variable occurs. It is formally defined
as below.

\begin{definition} 
Let $\A$ be the set of relational algebra formulae over
$\D$. Let $\varphi_1, \varphi_2 \in \A$ then all of the following
 also belong to $\A$:
(1) (selection) $\sigma_{i = x} \varphi_1$ where $i \in N$ and
	$x \in \vec{V} \cup \dom$; 
(2) (projection)
	$\pi_{i_1,i_2,...,i_k} \varphi_1$ where $i_1,i_2,...,i_k \in N$; 
(3) (cross-product)
	$\varphi_1 \times \varphi_2$; 
(4) (union)
	$\varphi_1 \cup \varphi_2$;  and
(5) (difference)
	$\varphi_1 - \varphi_2$. 
% When the difference operator is not included in a relational
% algebra formula $\phi$, we call it an SPCU formula.
\end{definition}

We now list one sample path transducer for each of the five
servlets.  We first describe the function 
of a servlet and then formally define it as a transition rule.
Figure\ \ref{fig:notations} summarizes the semantics
of all notations to be used in the formulae.

\begin{figure}
\begin{center}
\begin{scriptsize}
\begin{tabular}{|l|l|l||l|l|l|}
\hline 
 \tb{Var} & \tb{Type} & \tb{Meaning} & 
 \tb{Var} & \tb{Type} & \tb{Meaning} \\
\hline 
 $\#u$ & \ttt{Session Var} & \ttt{User Name}  &
 $\$s_I$ & \ttt{HTTP Param} & \ttt{session name for Insertsessions.php} \\
\hline 
 $\$s_A$ & \ttt{HTTP Param} & \ttt{session name for Addmember.php}  &
 $\$u_A$ & \ttt{HTTP Param} & \ttt{user name for Addmember.php} \\
\hline 
 $\$p_G$ & \ttt{HTTP Param} & \ttt{password for Generaloptions.php}  &
 $\$u_G$ & \ttt{HTTP Param} & \ttt{uname for Generaloptions.php} \\
\hline 
 $\$p_L$ & \ttt{HTTP Param} & \ttt{password for Login.php}  &
 $\$u_L$ & \ttt{HTTP Param} & \ttt{uname for Login.php} \\
\hline 
 $f$ & \ttt{HTTP Param} & \ttt{encryption function}  &
 $r$ & \ttt{HTTP Param} & \ttt{password safety constraint} \\
\hline 
 $U$ & \ttt{Relation} & \ttt{Users} & 
 $S$ &  \ttt{Relation} & \ttt{Sessions} \\
\hline 
 $M$ & \ttt{Relation} & \ttt{Members} & 
 &   &  \\
\hline 
\end{tabular}
\end{scriptsize}
\end{center}
\caption{Table of Notations}
\label{fig:notations}
\vspace*{-3mm}
\end{figure}
\vspace*{-2mm}

  \paragraph{Showsessions.php} The servlet first checks 
  the existence of a session variable ``\ttt{uname}" (represented
  by $\#u$), and then verifies if there is at least
  one session of which $\#u$ is a member. If the condition
  evaluates to \ttt{true}, the servlet executes a loop
  that displays the session name by retrieving information
  from table \ttt{Sessions} (\ttt{S}). Because the loop does not
  have any side effects (e.g., updating databases), the operations
  on \ttt{S} are not included in the path condition 
  during symbolic execution. The following constraint is used
  to represent the transducer.
  \begin{equation} \label{eq:s}
	\#u \neq \null \And \SAT(\sigma_{2=\#u} M)
  \end{equation}
  Here \SAT  tests if the input (a relational algebra formula)
  is satisfiable.  For convenience, we omitted the formula 
  $\#u'=\#u \And S'=S \And M'=M \And U'=U$ (i.e., maintaining
  the values of all variables and relations), but it needs to be encoded 
  in implementation.

 \paragraph{Insertsessions.php} The servlet inserts a new session record
  into relation $\ttt{S}$. It accepts one input parameter $\$s_I$ (the
  session name). The primary key \ttt{sid} is automatically incremented
  by the system. The action is modeled using the standard set union operator.
  It is also an example of integer constraints involved in the application.
  \begin{equation} \label{eq:i}
	S' = S \cup \{(|S|+1, \$s_I)\}
  \end{equation}

 \paragraph{Addmember.php} The servlet takes two parameters: user name
  ($\$u_A$) and session name ($\$s_A$). Then it looks up for
 the corresponding session id, and then inserts a tuple 
 into relation $\ttt{M}$ (\ttt{Membership}). Here we assume that the
 prime key constraint is valid
 (every record in \ttt{S} has a  distinct \ttt{sid}). So at any time,
 at most one new record is inserted into \ttt{M}.
  \begin{equation} \label{eq:a}
	M' = M \cup \pi_1(\sigma_{2=\$s_A}(S)) \times \{(\$u_A)\}
  \end{equation}

 \paragraph{Generaloptions.php} The transducer 
adds a user to the system.  It takes two parameters: 
user name ($\$u_G$) and password ($\$p_G$).
$f$ is an uninterpreted function that represents
the encryption procedure.
All uninterpreted functions are assumed to be deterministic. 
  \begin{equation} \label{eq:g}
	U' = U \cup \{(\$u_G, f(\$p_G))\} \And r(\$p_G)
  \end{equation}

In practice, there are security conditions that a legitimate user 
name and password must meet. This is represented using function $r(\$p_G)$ where
$r(x)$ is a Boolean function 
$\Sigma^* \rightarrow \{T,F\}$.
For example, we can define it as $r(x) = x.contains(``\#") \And |x|>7$.
The solution of string constraints can be
handled by string analysis (see e.g., \cite{nfm2010}).

 \paragraph{Login.php} The servlet takes two parameters: uname ($\$u_L$) and
password ($\$p_L$), and then verifies their existence in the database. It 
updates the session variable $\#u$ with the value of 
$\$u_L$, so that the user session can now be traced by $\#u$. 
The servlet assumes that $\#u$ is not set before 
the login activity. Also note that $(\$u_L,f(\$p_L)) \in U$
is a syntactic sugar for $\SAT(\sigma_{1=\$u_L}\sigma_{2=f(\$p_L)}(U))$.
  \begin{equation} \label{eq:l}
	(\$u_L,f(\$p_L)) \in U \And \#u' = \$u_L \And \#u=\null
  \end{equation}

\subsection{Discussions}
 We outline the idea of extracting path transducer models, though
 an implementation does not exist yet.
 Symbolic execution \cite{King76} is the major
 technical approach, and Halfond's recent work \cite{Halfond09issta}
 in interface extraction can be leveraged.
 A servlet is treated as a program, which takes global input (the GET and
 POST variables). These global variables are replaced by symbolic literals.
 When an assignment occurs, the variable on LHS (left hand side)
 is associated with a symbolic expression.
 Then, at a branch decision, the jump to one of the branches is made 
 randomly (so that it guarantees that each branch will be covered). 
 When entering a branch, the corresponding condition is 
 joined with the current {\em path condition}. 
 Clearly, the path condition records the conjunction of all conditions 
 that the initial input has to meet to reach the current execution location. 
 Some known system calls (e.g., those to backend databases) are intercepted
 and replaced with the corresponding symbolic expression. Others (unhandled)
 are abstracted as uninterpreted functions. Eventually when the
 symbolic execution completes,
 the path transducer is the conjunction of three components: (1) the
 path condition, (2) assignments that change values of
 global session variables, and
 (3) operations that update the contents of database relations.
 String analysis plays an important role in deciding the precision
 of the translation from symbolic string expressions to
 relational algebra formulae. Given a string expression (consisting of
 constant words and string variables) that represents a SQL query,
 populating the string variables with ``benign" values generates
 a real SQL query which can be parsed to a SQL syntax tree and
 then be translated into relational algebra formulae (with
 variables reloaded). Note that here we ignore SQL injection attacks.

\section{Relational Constraint}

 This section introduces the preliminary results about relational constraints.
% The general satisfiability problem on relational algebra is undecidable, 
%as shown by the similar results 
%on relational calculus in \cite{AHV95}. 
Inspired by the work on testing DBMS by Khurshid {\it et al.}
\cite{ADUSA}, we translate a relational algebra formula  to a relational logic
formula and then use Alloy Analyzer \cite{Alloy} to find a model
in a finite scope.  Our initial experiments show that
primary and foreign key constraints can be conveniently modeled.
Within a small scope, Alloy can quickly find solutions that
satisfy a relational constraint.

Let $\caD$, $\vec{V}$, $\dom$ represent the data schema, set of session
variables, and the data domain, respectively.
Given a transducer
$T=(\caD, \dom, \vec{V}, \vec{A})$, we say a state of
$T$ is a database instance of $\caD$ and a valuation of
all variables in $\vec{V}$. Given two states $s$ and $s'$,
we write $s' \in T(s)$ if $s'$ is the result of
applying all the assignments of $T$ on $s$, i.e., $T(s,s')$
evaluates to true.  Note that there might be multiple
states resulting from the same transition and input state.
If $s'$ is the only such post-state, we write $s' = T(s)$.
We now define the notion of symbolic image computation.

\begin{definition}
  Let $T=(\caD,\dom,\vec{V},\vec{A})$ be a path transducer 
and $I$ is a relational constraint over $\caD$ and $\vec{V}$.
The preimage $\Pre(T,I)$ is defined as
$\Pre(T,I) = \{s ~|~ s' \in T(s) \And s' \in I\}$, and
the postimage is $\Post(T,I) = \{s' ~|~ s' \in T(s) \And s \in I\}$.
When $I$ is \ttt{true}, we simply write
$\Pre(T,I)$ and  $\Post(T,I)$ as $\Pre(T)$ and $\Post(T)$, respectively.
\end{definition}

The complexity of generating a relational constraint
that represents $\Pre(T,S)$ is in polynomial time.
This can be easily inferred from the fact that
a transducer $T$ is a collection of assignments on relations and
variables. Simply replace all primed variable and relation
by its RHS. Note that we do not have a similar result for
post-image. We now proceed to a finite scope solution
for the satisfiability problem. We illustrate the idea using
an example.

\begin{figure}[h!]
\begin{minipage}[b]{0.4\linewidth}
\centering
\lstset{language=PHP, frame=single,  
label=listing:show, basicstyle=\scriptsize, captionpos=b}
\begin{lstlisting}
1  sig vchar {}
2  sig UserRecord{
3   uname: vchar,
4   pwd: vchar
5  }
6
7  sig SessionRecord{
8   sid: Int,
9   sname: vchar
10 }
11
12 sig MemberRecord{
13  sid: Int,
14  uname: vchar
15 }
16
17 sig UserTable{
18  list: set UserRecord
19 }{
20  //primary key
21  all x,y: list | 
22    x.uname = y.uname => x=y
23 }
24
25 sig SessionTable{
26  list: set SessionRecord
27 }{
28  //primary key
29  all x,y: list | 
30   x.sid = y.sid => x=y
31 }

\end{lstlisting}
\end{minipage}
\hspace{0.5cm}
\begin{minipage}[b]{0.5\linewidth}
\centering
\lstset{language=PHP, frame=single,  
label=listing:show2, basicstyle=\scriptsize, captionpos=b}
\begin{lstlisting}
32 sig MemberTable{
33   list: set MemberRecord
34 }{
35  //primary key ...  and foreign key 1...
36  //foreign key 2
37  all x: list | one y: SessionTable | 
38   one z : y.list | z.sid = x.sid
39 }
40 
41 fact aboutRecords{
42   ( all x: UserRecord | 
43      one u: UserTable | x in u.list )
44   ...
45 }
46
47 // The following are for query
48 one sig c_s1, c_s2 extends vchar {}
49 pred part1[d:vchar]{
50  some a,c:Int | some b: vchar | 
51   a=c && b in c_s1 && 
52   (some y: SessionTable | some x: y.list | 
53      x.sid=a && x.sname=b) 
54   && (some y: MemberTable | some x: y.list | 
55      x.sid=c && x.uname=d)
56 }
57 pred part2[d:vchar]{
     ...
58 }
59 pred query[d:vchar]{
60   part1[d] and !part2[d]
61 }

\end{lstlisting}
\end{minipage}
  \caption{Sample ALLOY Specification}
  \label{fig:alloy}
\end{figure}

\begin{examp} Assume that we are interested in performing the following 
query: list all users who are members of session ``s1" but
not of session ``s2". This query can be expressed using
the following relational algebra formula:
\[
	\pi_4(\sigma_{2='s1'} \sigma_{1=3} (S \times M)) -
	\pi_4(\sigma_{2='s2'} \sigma_{1=3} (S \times M))
\]
The Alloy specification of the query is shown in Figure\ \ref{fig:alloy}.
The first part (lines 1 to 40) defines the data schema: 
relations and the corresponding primary key constraint for
each relation. Each row of a relation is defined as
a \ttt{sig} (data type) in Alloy, and each column is defined
as an attribute. A relation is defined as a set of the corresponding
rows (records). Constraints (such as primary and foreign keys)
can be defined conveniently as first order relational logic formula.
The \ttt{aboutRecords} fact asks Alloy to perform
search on those records in a relation only.

The second part of the Alloy specification encodes the 
query. It essentially follows the idea of converting a relational
algebra to a first order logic query. The predicate
\ttt{query} encodes the desired difference operation.
By running the query for exactly 1 instance of each
data relation and 3 instances of data records for each relation,
a database instance is generated in 78ms to satisfy the query, on
a laptop PC with 4GM memory and 3GHz CPU.
It is shown as below:
\begin{enumerate}
\setlength{\itemsep}{0.2pt}
\item \ttt{SessionTable: \{($vchar3$,$c\_s1$)\}}
\item \ttt{MemberTable: \{($vchar3$,$vchar2$)\}}
\item \ttt{UserTable: \{($vchar2$,$vchar3$)\}}
\end{enumerate}

Here all foreign key constraints are followed, e.g.,
the \ttt{sid} column (\ttt{vchar3}) of the only record of
\ttt{MemberTable} is the same as the one in
\ttt{SessionTable}. $c\_s1$ denotes the constant
$s1$, and it is the value of the \ttt{sname} of
the \ttt{SessionRecord}. For simplicity, the uninterpreted function $f$ (encryption) is 
not encoded in this example. It is available in the Alloy model for
Step 5 in \S{5}.
\end{examp}

\section{Call Sequence Synthesis}
\def \h[#1]{{\vec{h}(#1)}}
\def \p[#1]{{\vec{p}(#1)}}
This section introduces the synthesis algorithm that
generates a test case composed of a sequence of
HTTP requests. The purpose is to reach a certain
database state for achieving coverage goals.
We begin with some formal definitions needed for
discussions later.

 \begin{definition}An HTTP request $r$ is a tuple
$(u, \vec{p})$ where $u$ is the base URL, and
$\vec{p}$ is a finite set of parameters. 
Each parameter is a tuple $(n,v)$ where $n$ is the name
of the parameter and $v$ is its value. $\vec{p}[n]$ denotes the value
of parameter $n$.
 \end{definition}

 \begin{definition} A call sequence is a finite sequence of HTTP requests
 $r_1, ..., r_n$. We call the corresponding HTTP responses 
 $s_1, ..., s_n$. Each $s_i$ is a string that represents
 the contents of the HTML file returned.
 \end{definition}
 A call sequence is also written as $(r_1,s_1),(r_2,s_2),...,(r_n,s_n)$ when
 HTTP responses need to be considered.
 We call a response {\em bad}, if it contains 
 error messages such as \ttt{HTTP 505 internal error}. 
% We could also
%  define a web servlet from a perspective of functions.
%  Let $\caR$ and $\caS$ be the domain
% of all HTTP requests and HTTP responses. 
% A web application servlet $t: \caR \rightarrow \caS$
% is a {\em deterministic} function, which given a HTTP request $r$, 
% returns a HTTP response $s$. 
% Given a servlet $t$, let $T(t)$ be the set of path transducers induced from
%   $t$. $T(t)$ could be an infinite set. Let $T(t,n)$ be a subset
% with $n$ distinct elements.

 \subsection{Call Sequence Synthesis Algorithm}
 Figure\ \ref{fig:callseq} presents a heuristic algorithm that
synthesizes a call sequence of web servlets. Its input includes
(1) initial database states specified using a relational constraint
$S_0$, (2) a finite set of transducers $\caT$, and (3) 
the objective states represented using relational constraint
$\varphi$.

 As shown in Figure\ \ref{fig:callseq}, the backtracking algorithm
attempts to simulate the changes of system states, using
pre-image computations. Here $S$ is a stack which stores
the call sequence. Each element of $S$ is a tuple which
records the current servlet being attempted and the corresponding
system states (represented using a relational constraint).
Every iteration of the backtracking loop tries to identify
a new servlet which modifies the system state, geared towards
the direction of initial states $S_0$. The current system states
(represented using $\varphi$) are compared with $S_0$ using
function \ttt{getModified}. It tries to extract the set of
variables and relations whose ``value domain"
change between the two sets of symbolic states. 
The value domain of a variable $v$ in constraint $\varphi_1$
is formally defined using formula
$\varphi_1(v)=\{a ~|~ s \in \varphi_1 \And s[v]=a\}$. 
Here $s[v]$ is the value of variable $v$ in state $s$.
Clearly $\varphi_1(v)$ can be computed using existential elimination
of all other variables/relations in the formula of $\varphi_1$.
If a finite scope is given, this can be achieved using
first order relational logic in Alloy \cite{Alloy}.
Once the ``difference" of the two symbolic constraints
$\varphi_1$ and $\varphi_2$ is found, each servlet
is statically examined. A servlet that modifies some
variables/relations in $M_1-M_2$ is identified and pushed to
stack $S$. If none is found, the search process backtracks.
It proceeds until the initial state constraint $S_0$
has some intersection with the current system state; or
the algorithm fails and returns an empty stack.

  We use a case study to illustrate the effectiveness of
the algorithm. Applying the algorithm to real-world examples
remains as our future work. The heuristics works better
with \ttt{INSERT} statements than \ttt{UPDATE} statements. 
 
 \begin{figure}[h!]
 \begin{scriptsize}
 \begin{tabbing}
 ~~ \= ~~ \= ~~ \= ~~ \= ~~ \= ~~ \= ~~ \= ~~ \= ~~ \= ~~ \= ~~ \= ~~  \\
1  //$S_0$: desired initial states,  $\varphi$: desired path condition,
  $\caT$: a finite set of path transducers \\
2  Procedure CallSeqGen($\caT$, $S_0$, $\varphi$)  \\
3  \> \> Let $S = [(\bot,\varphi)]$  \\
4  \> \> //$S$ is a stack which stores the intended call sequence to return \\
5  \> \> //Each element of $S$ is a tuple of (action, current state) \\
6  \> \> while($S_0 \intersect \varphi$=$\emptyset$ and $|S|\geq 1$)\{ \\
7  \> \> \> Let $\varphi_1$ and $\varphi_2$ be the states stored in the top 2 
	elements of $S$.
	If $|S|=1$, let $\varphi_2$ be \ttt{true} \\
8  \> \> \> Let $M_1$ = getModified($\varphi_1$,$S_0$) \\
9  \> \> \> Let $M_2$ = getModified($\varphi_2$,$S_0$) \\
10  \> \> \> Find a path transducer $s \in \caT$ which modifies some
target $v \in M_1-M_2$ and $\Pre(s,\varphi) \neq \ttt{false}$\\ 
11  \> \> \> If no servlet is available then backtrack: \\
12  \> \> \> \> remove the first element of $S$; continue \\
13  \> \> \> $\varphi := \Pre(s, \varphi)$; $S := (s,\varphi) \circ S$ \\
14  \> \> \} \\
15  \> \> return $S$\\
  \\ 
16  //return a set of session variables and relations that are changed
	between the two sets of states represented 
	by $\varphi_1$ and $\varphi_2$ \\
17  Procedure getModified($\varphi_1$,$\varphi_2$)  \\
18  \> \> Let $R = \{\}$  // $R$ is the set of variables and relations to return\\
19  \> \> foreach $v \in V \cup \caD$: //$v$ could be a session variable or relation\\
20  \> \> \> Let $\varphi_1(v) = \{a ~|~ s \in \varphi_1 \And s[v]=a \}$ 
	//here $s$ is a state belong to $\varphi_1$ and $s[v]$ is the value
	of $v$ in $s$ \\
21  \> \> \> Let $\varphi_2(v) = \{a ~|~ s \in \varphi_2 \And s[v]=a \}$ \\
22  \> \> \> if $\varphi_1(v)-\varphi_2(v) \neq \emptyset$ or
		$\varphi_2(v) - \varphi_1(v) \neq \emptyset$:  $R = R + \{v\}$ \\
23  \> \> return $R$\\
 \end{tabbing}
 \end{scriptsize}
 \vspace*{-3mm}
 \caption{Call Sequence Generation Algorithm}
 \label{fig:callseq}
 \end{figure}
 \vspace*{-3mm}

%  \begin{thm} The \ttt{CallSeqGen} algorithm shown in Fig. 2 can always 
% terminate. The worst complexity is PSPACE. 
%  \end{thm}
% 
%  Proof sketch: 
%  Let $|W|$ represent the sum of the size of database schema and the set
%  of global session variables. Clearly, $|W|$ is finite. The number of iterations
%  of the $\ttt{while}$ loop in Fig.\ \ref{fig:callseq} is at most
%  $|W|^2$ as every iteration tries to set one session var/data table from
%  the precondition state. The solution and satisfiability of $\Pre$ constraint
%  is PSPACE, which leads to the conclusion. 

\subsection{Case Study}

The coverage goal is to reach line 45 in Listing 1. We would like to
synthesize a call sequence that reaches the line. 
The initial state $S_0$ is expressed using formula  
$\#u=\null \And S=M=U=\emptyset$, i.e., 
the database is empty and none of the global session variables is set.
The target constraint $c_0$ is represented by
relational constraint
$\#u\neq\null \And \SAT(\sigma_{2=\#u} M)$, i.e., table $M$ has
one record whose second attribute ``\ttt{uname}" has
the value $\#u$ (the user name contained in session variable \ttt{\#uname}).
This is essentially the pre-condition of
 a path transducer of \ttt{Showsessions.php} that reaches line 45.

\paragraph{Step 1} We start with the path transducer of
\ttt{Showsessions.php}. The first step is to
compute the pre-image of the transducer.
\begin{scriptsize}
\begin{equation}
\Pre(\#u\neq\null \And \SAT(\sigma_{2=\#u} M) \And 
	S'=S \And U'=U \And M'=M \And \#u'=\#u 
     ,\ttt{true})
\end{equation}
\end{scriptsize}
 Recall that $\Pre$ takes two parameters: (1) the transition relation, and
(2) the post image.  The algorithm of $\Pre$ is straightforward. First,
convert all ``current state" variables in the post condition (i.e.,
the second parameter of $\Pre$) to ``next state", construct the
conjunction of the two
parameters, and then replace 
each post state variable (e.g.,``$S'$") with its RHS (e.g., ``$S$")
in the assignment of transition relation. 
We then obtain the precondition. Notice that this algorithm
is based on the assumption that in  each assignment, post-state variables
occur in LHS only.
%(2) only equality comparison is involved (we do not
%support any arithmetic or string operator at this moment).
  Clearly, for step 1, the resulting pre-image is shown below:

\begin{scriptsize}
  \[
	N_4: \#u\neq\null \And \SAT(\sigma_{2=\#u} M)
  \]
\end{scriptsize}
\vspace*{-3mm}

  The synthesis algorithm cannot terminate here, because the goal
$S_0 \intersect N_4 \neq \emptyset$ is not met. 
% Note that here we need algorithms for
% deciding the containment of relational constraint, which is
% discussed in the previous section. 
% Now, the synthesis heuristic steps in. 
We need to determine which is
 the action before \ttt{Showsessions.php}. 
% The heuristics is
% to statically analyze each of the transducers and select them
% which affect the session variables and relations occurring in $N_4$.
The candidates are 
\ttt{Login.php} which resets $\#u$ and \ttt{Addmember.php}
which updates $M$. Without loss of generality, we choose \ttt{Login.php}
as the preceding action of \ttt{Showsessions.php}.

\paragraph{Step 2} The current call sequence is \ttt{LS} (\ttt{Login.php}
and then \ttt{Showsessions.php}). The post condition of \ttt{L} is $N_{4}$.
We can compute its pre-image as follows.

\begin{scriptsize}
\begin{eqnarray*}
 &\Pre(&(\$u_L,f(\$p_L)) \in U \And \#u=\null \And \#u'=\$u_L \And 
 	U' = U \And S'=S \And  M'=M \And \\
 & &	\#u'\neq\null \And \SAT(\sigma_{2=\#u'} M'))
\end{eqnarray*}
\end{scriptsize}
Here the first row is the transition rule for \ttt{Login.php},
and the second row is the primed form of $N_4$. We use the
one parameter version of the $\Pre$ function here, and
the post-image has to be primed.  The same algorithm results in the 
following pre-image. Note that here
$\$u_L$ and $\$p_L$ are the input parameters (uname and pwd)
of \ttt{Login.php}.

\begin{scriptsize}
\[
 N_3: (\$u_L,f(\$p_L)) \in U \And \#u=\null \And \$u_L\neq\null
	\And \SAT(\sigma_{2=\$u_L} M)
\]
\end{scriptsize}
\vspace*{-3mm}

\paragraph{Step 3} Next, 
\ttt{Addmember.php} is identified by the heuristic algorithm as the
preceding action. The pre-image is computed as below.
As usual, the first row is the transition rule and the
second row is the primed form of $N_3$. Notice that
$\$u_L$ is the input parameter for \ttt{Login.php}.
It is treated as a symbolic literal (its value cannot change
during the execution). We do not have to ``prime" it in
the post-image.

\begin{scriptsize}
\begin{eqnarray*}
 &\Pre(& M'=M \cup (\pi_1(\sigma_{2=\$s_A}(S)) \times \{(\$u_A)\})
	\And 
 	U' = U \And S'=S \And  \#u'=\#u \And \\
 & &       (\$u_L,f(\$p_L)) \in U' \And \#u'=\null \And \$u_L\neq\null
	\And \SAT(\sigma_{2=\$u_L} M') )
\end{eqnarray*}
\end{scriptsize}
\vspace*{-3mm}

Here $\$u_A$ and $\$s_A$ are the input parameters 
(uname and session name) of the \ttt{Addmember.php}.
The pre-image computation results in the following state constraint:

\begin{scriptsize}
 \[
	\SAT(\sigma_{2=\$u_L}(M ~\cup~ (\pi_1(\sigma_{2=\$s_A}(S))) \times \{(\$u_A)\}))
	\And (\$u_L,f(\$p_L)) \in U \And \#u=\null
	\And \$u_L\neq\null 
 \]
\end{scriptsize}
\vspace*{-3mm}

\paragraph{Step 4} The pre-image computation is listed as below
for \ttt{Generaloptions.php} (to add a user). Here $\$u_G$ and
$\$p_G$ are the input parameters (uname and pwd).

\begin{scriptsize}
\begin{eqnarray*}
 &\Pre(&  U'=U \cup \{(\$u_G,f(\$p_G))\} \And r(\$p_G) \And 
 	M' = M \And S'=S \And  \#u'=\#u \And \\
   & &  	\SAT(\sigma_{2=\$u_L}(M' ~\cup~ (\pi_1(\sigma_{2=\$s_A}(S'))) \times \{(\$u_A)\}))
	\And (\$u_L,f(\$p_L)) \in U' \And \#u'=\null
	\And \$u_L\neq\null)\\
\end{eqnarray*}
\end{scriptsize}
\vspace*{-5mm}

It results in the following:
\begin{scriptsize}
\[
 N_1: r(\$p_G) \And 
	\SAT(\sigma_{2=\$u_L}(M~\cup~ (\pi_1(\sigma_{2=\$s_A}(S))) \times \{(\$u_A)\}))
	\And (\$u_L,f(\$p_L)) \in U \cup \{(\$u_G,f(\$p_G))\}
	\And \#u=\null \And  \$u_L\neq\null 
\]
\end{scriptsize} 

Since $S_0 \intersect N_1 =  \emptyset$,
The heuristic algorithm has to proceed to modify
table \ttt{S}.

\paragraph{Step 5} Using constraint $N_1$ as the post condition of
\ttt{Insertsession.php}, we have: 

\begin{scriptsize}
\begin{eqnarray*}
 &\Pre(&  S' = S \cup \{(|S|+1, \$s_I)\} \And 
 	M' = M \And U'=U \And  \#u'=\#u \And \\
 & & 
 r(\$p_G) \And 
	\SAT(\sigma_{2=\$u_L}(M' ~\cup~ 
		(\pi_1(\sigma_{2=\$s_A}(S'))) \times \{(\$u_A)\}))
	\And (\$u_L,f(\$p_L)) \in U' \cup \{(\$u_G,f(\$p_G))\}
	\And \#u'=\null \And  \$u_L\neq\null)
\end{eqnarray*}
\end{scriptsize}

This leads to the final constraint $N_0:$ 
\begin{scriptsize}
\[
 N_0: r(\$p_G) 
	\And \SAT(\sigma_{2=\$u_L}(M \cup (\pi_1(\sigma_{2=\$s_A}(S \cup \{(|S|+1,\$s_I)\})))
		\times \{\$u_A\}))
	\And (\$u_L,f(\$p_L)) \in U \cup \{(\$u_G,f(\$p_G))\}
	\And \#u=\null \And \$u_L\neq\null
\]
\end{scriptsize} 

If we test $S_0 \intersect N_0$, the following is
an assignment which provides satisfiability:
\begin{scriptsize}
\[
	M=S=U=\emptyset \And
	\$u_L=\$u_G=\$u_A=a \And
	\$p_G=\$p_L=b \And
	\$s_I=\$s_A=c \And
	\#u=\null
\]
\end{scriptsize}
Here $a$,$b$,$c$ are three constants generated by the model finder.
The constraint $r(b)$ can be discharged separately using
a string constraint solver like \cite{Yu08spin} and \cite{nfm2010}.
When the string constraint $r(b)$ is ignored, Alloy Analyzer is
able to generate the model in 1.07 seconds and solve the model
in in 57ms. In the model generated by Alloy, constants $a$ is equal to $c$,
and the encryption function is properly  modeled in Alloy and
it has the property: $\forall a,b: f(a)=f(b) \Leftrightarrow a=b$.

% Our analysis eventually leads to a test case (call sequence) generated,
% as specified in Figure\ \ref{fig:goal}.

% \begin{figure*}[t!] 
% \begin{center}
% \begin{scriptsize}
% \begin{tabular}{|c|c|c|c|}
%  \hline  
%  \tb{Step} & \tb{Servlet} & \tb{DB State Before Invocation} & \tb{Input} \\
%  \hline  
%  1 & \ttt{Insertsessions.php} & $S=M=U=\emptyset$ & $\$s_I: \ttt{'a'}$ \\
%  2 & \ttt{Generaloptions.php} & $S=\{(1,\ttt{'a'})\} \And M=U=\emptyset$ & $
% \$u_G:\ttt{'a'} \And \$p_G:\ttt{'\#aaaa'}$ \\
%  3 & \ttt{Addmembers.php} & 
% $S=\{1,'a')\} \And U=\{(\ttt{'a','\#aaaa')}\} \And M=\emptyset$ 
% & $\$u_A: \ttt{'a'} \And \$s_A: \ttt{'a'}$ \\
%  4 & \ttt{Login.php} &  
% 	$S=\{1,\ttt{'a'})\} \And U=\{(\ttt{'a','\#aaaa'})\} \And M=\emptyset$ 
% 	& $\$u_L: \ttt{'a'} \And \$p_L: \ttt{'a'}$ \\
%  5 & \ttt{Showsession.php} & 
% 	$S=\{1,\ttt{'a'})\} \And U=\{(\ttt{'a','\#aaaa'})\} \And 
% 		M=\{(1,\ttt{'a'})\} \And \#u=\ttt{'a'}$ 
% 	&   \\
%  \hline  
% \end{tabular}
% \end{scriptsize}
% \end{center}
% \caption{Test Sequence}
% \label{fig:goal}
% \end{figure*}

\section{Detecting Workflow Attack}
%1. introduction
  This section briefly discusses the extension of the algorithm
for detecting workflow attacks \cite{Balzarotti07CCS}. 
% Different from desktop software
% systems, web applications adopt an open architecture -- 
Since web servlets are openly accessible to end users,
hackers could potentially access the web application and
violate its intended ``workflow" logic (e.g., shipping
a product before payment is handled).
Such an attack can cause great financial losses.
  To model the ``intended" workflow of a web application, we could
apply string analysis like \cite{Christensen03SAS} and collects the URLs 
that are contained in the HTML generated by a servlet. Then, a workflow
attack can be defined formally.

\begin{definition} An enhanced path transducer is a tuple
$T=(\caD, \dom, \vec{V}, \vec{A}, \vec{L}, U)$ where
$\caD$, $\dom$, $\vec{V}$, $\vec{A}$ are as defined
in Definition\ \ref{deff:transducer}. $\vec{L}$ is
a set of transducers that $T$ can navigate to. $U$ is
the base URL where the corresponding servlet is deployed.
\end{definition}

Given a collection of path transducers, we use $\vec{L}(U)$
to denote the union of the $\vec{L}$ components of all
path transducers who have $U$ as the base URL.

\begin{definition} Given a web application that consists of $\caT$, 
a finite set of path transducers, a workflow attack is
a call sequence $(r_1,s_1), ..., (r_n,s_n)$ where
none of the responses is bad
and there exists $i \in [1,n-1]$ s.t.
$r_{i+1}[0] \not\in \vec{L}(r_i[0])$. Here 
$r[0]$ to refer to the first element, i.e., the request URL, of a request $r$.
\end{definition}

%5. discussion of detection algorithm
  The \ttt{CallSeqGen} algorithm in Figure\ \ref{fig:callseq} can
be slightly modified to discover workflow attacks. The inputs
to the algorithm are: (1) $\caT$, the collection of enhanced  
path transducers; (2) $S_0: \forall v \in V~ v=\null$, the initial
state where all session variables are $\null$ (database is not
necessary to be empty); and (3) \ttt{true} as the desired 
post condition. 
Then at line 10 of
Figure\ \ref{fig:callseq}, append an additional condition shown as
below when selecting $s$: 
\[
  |S|=2 \Rightarrow s' \not\in \vec{L}(s) 
\] 
where $s$ is the path transducer to be selected, $s'$ is the
second top element currently in the stack, and $\vec{L}(s)$ is the
$\vec{L}$ component of $s$. Enforcing $|S|=2$ is to find the
shortest attack string, such that the violation action is the last one
in the sequence. Note that due to the backtracking nature of the
algorithm, if the current guess of $s$ does not work, the algorithm
will trace back to find another candidate.

\section{Related Work and Conclusion}
%  We have outlined a framework for automatically generating
%test cases of web applications. The key idea is to first extract
%the interface of each web servlet as a set of single path
%relational transducers. 
% Then, by taking advantage of Alloy Analyzer,
%, which allows us to reach a certain database
% state that is required for achieving a code coverage goal.
% Our future work includes the completion of the implementation,
% and investigating more efficient solving techniques for
% symbolic relational constraints.

  %1. database testing
  This work is closely related to 
testing database applications,
e.g., code coverage and database
unit testing \cite{Agenda05icse,Emmi07issta,Bruno05VLDB}. 
Recently, query aware database generators are reported to
significantly improve the size and quality of test cases
(see \cite{ADUSA,Lo07icde}).  Our work goes one more step beyond
database generation -- we synthesize the call sequence of web application
servlets.

  %The query aware database generation problem, i.e., to generate
%a database instance that satisfies a query, subsumes the query 
%satisfiability problem of first order logic \cite{AHV95}. Hence, 
The general satisfiability problem for relational database
is undecidable \cite{AHV95}. In practice, one has to adopt
either approximation or solving decidable fragments.
Emmi, Majumdar, and Sen generate database input for programs
in \cite{Emmi07issta}, and a decidable  fragment
of SQL is handled, which does not allow join and negation.
In \cite{Lo07icde}, a reverse query processing technique is developed,
which introduces approximation when handling negation. 
In \cite{ADUSA}, Khalek {\it et al.}
translate SQL queries to relational logic formula, and
use Alloy Analyzer \cite{Alloy} to perform model finding.
We adopt a similar approach to that of Khalek's.
% despite a slight syntactic difference in translation.
% Later, the efficiency can be improved by
% calling the Kodkod model finder \cite{Kodkod} directly.

  %2. relational transducer 
  Extracting the interface of a web servlet has been
investigated by Halfond and Orso 
\cite{halfond07fse,halfond08fse},
where static analysis is used to identify the parameters
accepted by a web application. In \cite{Halfond09issta},
they went a step further to collect path conditions as 
web servlet interfaces.
In this paper, the interface extraction concentrates
on the manipulation of backend databases.
Each web application servlet is modeled as a set of
path transducers, which is
inspired by the relational transducer model \cite{AVFY00}
introduced by Abiteboul {\it et al.} for modeling electronic commerce.
Automated verification of relational transducers
is discussed in \cite{Spielmann03,Deutsch04}. 
The problem is in general undecidable.  
% Some decidable fragments, e.g., 
% by bounding occurrence of free variables, are shown by Deutsch
% in \cite{Deutsch08}. The synthesis algorithm presented in this paper
% can be regarded as a lighter weight approach than \cite{Spielmann03,Deutsch04,Deutsch08}, as there is no need to handle temporal logic.
% However, symbolic image computation (to compute pre/post conditions)
% is needed for the algorithm. 

  %4. web application testing
  Testing web applications has its unique challenges. Due to the
existence of server states, e.g., session variables and
HTTP cookies, a test case of web applications usually consists
of a sequence of HTTP requests. This is often called 
session based testing \cite{Sampath03,Elbaum05tse,Sprenkle05ase}.
In the aforementioned work, HTTP sessions are either manually
created or collected by parsing Apache server log. The
technique presented in this paper synthesizes test cases automatically.
It has the potential to improve code coverage.

  We have outlined a framework for automatically generating
test cases of web applications. The key idea is to first extract
the interface of each web servlet as a set of single path
relational transducers.  
Then we could solve symbolic constraints on relational databases
and synthesize a call sequence of
web servlets.  Our future work includes
the implementation of the proposed technique and investigation
of more efficient constraint solving techniques.

\bibliographystyle{plain}
\bibliography{fu}

\end{document}